\documentclass[twoside]{ilcws10}
\usepackage[latin1]{inputenc}
\usepackage[dvips]{graphicx,epsfig,color}
\usepackage{wrapfig,rotating}
\usepackage{amssymb,amsmath,array}
\usepackage{hyperref}

\begin{document}
\title{
%%%%   Paper title goes here  %%%%%%%%%%%%%%
Shintake Monitor in ATF2 : Present Status} %% 
%***********************************************************************
% AUTHORS INFORMATION AREA
%***********************************************************************
\author{T.~Yamanaka$^1$, M.~Oroku$^1$, Y.~Yamaguchi$^1$, Y.~Kamiya$^2$,
T.~Suehara$^2$, S.~Komamiya$^1$,\\
T.~Okugi$^3$, N.~Terunuma$^3$, T.~Tauchi$^4$, S.~Araki$^3$, J.~Urakawa$^3$
% Optional short acknowledgment: remove next line if non-needed
\thanks{This is work is supported by KEK.}
% DO NOT MODIFY THE FOLLOWING '\vspace' ARGUMENT
\vspace{.3cm}\\
% Addresses and institutions (remove "1- " in case of a single institution)
1- The University of Tokyo - Department of Physics\\
7-3-1, Hongo, Bunkyo, Tokyo - Japan
%% Remove the next three lines in case of a single institution
\vspace{.1cm}\\
2 - The University of Tokyo - ICEPP\\
7-3-1, Hongo, Bunkyo, Tokyo - Japan
\vspace{.1cm}\\
3 - KEK - Accelerator Laboratory\\
1-1, Oho, Tsukuba, Ibaraki - Japan
\vspace{.1cm}\\
4 - KEK - Institute of Particle and Nuclear Science\\
1-1, Oho, Tsukuba, Ibaraki - Japan\\
}
%%***********************************************************************
% END OF AUTHORS INFORMATION AREA
%***********************************************************************

\maketitle

\begin{abstract}
A beam size monitor so called Shintake monitor, which uses the inverse Compton scattering between 
the laser interference fringe and the electron beam was designed for and installed at ATF2. 
The commissioning at ATF2 was started in the end of 2008 and succeeded in the measurement
 of the fringe pattern from the scattered gamma-rays. 
The present status of the Shintake monitor is described here.
\end{abstract}

\section{Overview}

\subsection{ATF2}
The Accelerator Test Facility (ATF) was constructed at KEK in Japan to
achieve small vertical emittance needed for ILC. The performance
of damping ring has been studied here. ATF2 was constructed as an extension 
of ATF to test the ILC beam delivery system including the final focusing scheme\cite{ATF2_proposal}.
By using small emittance beam provided from ATF damping ring, ATF2 aims to
focus the vertical beam size to 37 nm at the virtual interaction point.

\subsection{Shintake Monitor}
\begin{wrapfigure}{r}{0.5\columnwidth}
%\begin{figure}
\centerline{\includegraphics[width=0.45\columnwidth]{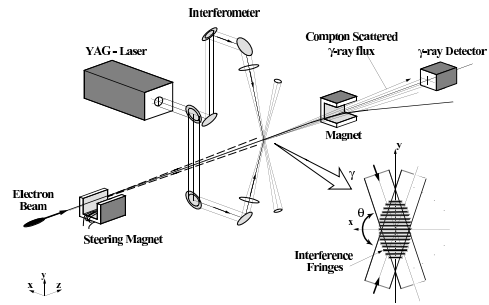}}
\caption{schematic of the Shintake monitor (original figure from\cite{shintake_preprint}) }\label{fig:shintake_schematic}
\end{wrapfigure}
%\end{figure}

Shintake monitor is a beam size monitor for the electron beam utilizing the laser interference
fringe pattern, which was originally designed by T~Shintake for the FFTB experiment
at SLAC\cite{shintake_NIM}.
Figure \ref{fig:shintake_schematic} shows the schematic of Shintake monitor.
The laser light is split and crossed at the interaction point of electron beam.
The electron beam passes through the interference fringe pattern formed
at the interaction point and the inverse Compton scattering occurs.
If the beam size is small relative to the fringe pitch, the number of scattered 
photons at the peak of fringe and that at the valley of the fringe differ significantly.
On the other hand, if the beam size is comparable to the fringe pitch, the electron
beam interact with the the peak and valley at the same time. Then
the numbers of scattered photons do not change so much.

Define the modulation depth $M$ as follows,
\begin{equation}
M = \frac{N_{\rm max} - N_{\rm min}}{N_{\rm max}+N_{\rm min}}
\end{equation}
where $N_{\rm max}$ is the measured maximum number of photons and 
$N_{\rm min}$ is the measured minimum number of photons when the 
relative position between the electron beam and the laser interference
fringe is moved. Then this modulation depth becomes the function of 
the electron beam size and can be expressed by the following equation.
\begin{equation}
M = |\cos \theta | \exp \left[ -2(k_y \sigma_y )^2 \right]
\label{eq:modulation_depth}
\end{equation}
where $\theta$ is the crossing angle between the two laser lights,
$k_y$ is the wave number of the interference fringe in vertical and
the $\sigma_y$ is the vertical electron beam size.

\section{Components of Shintake Monitor}
\begin{wraptable}{r}{0.5\columnwidth}
\centerline{\begin{tabular}{|l|r|}
\hline
wavelength & 532 nm \\ \hline
line width & $<$ 0.003 cm$^{-1}$ \\ \hline
repetition rate & 6.25 Hz \\ \hline
pulse width & 8 ns (FWHM) \\ \hline
timing jitter & $<$ 1 ns (RMS) \\ \hline
pulse energy & 1.4 J \\ \hline
\end{tabular}}
\caption{Specification of the laser.}
\label{tab:laser_spec}
\end{wraptable}

Shintake monitor is composed of mainly the following three parts. 
\begin{itemize}
\item Laser system
\item Laser optics
\item Gamma-ray Detector
\end{itemize}

\subsection{Laser System}

In Shintake monitor at ATF2, 2nd harmonics of a pulsed Nd:YAG laser (PRO350, SpectraPhysics) is used.
Specification of the laser is listed in Table \ref{tab:laser_spec}

\subsection{Laser Optics}
The purposes of the laser optics are 
\begin{enumerate}
\item transport the laser light to the beam line
\item split the laser light 
\item cross the laser lights at the interaction point of the electron beam
\item move the interference fringe
\end{enumerate}
The measurable beam size range of Shintake monitor is very limited because it cannot measure
the beam larger than the half of the fringe pitch. 
Therefore we prepare 3 types of the laser crossing angles to widen the measurable range\cite{kamiya_TILC}. 
The fringe pitch
can be changed by switching the laser crossing angles according to the current beam size. 

The position of the interference fringe is moved by changing the relative phase between
two split laser lights. The intensity of the laser interference fringe at the interaction point
can be written by
\begin{equation}
I(y) = I_0 \left( 1+ \cos \theta \cos (2k_y y + \alpha )\right)
\label{eq:laser_dist}
\end{equation}
where $y$ is the coordinate in the vertical direction,
definition of $\theta$ and $k_y$ are same as Equation (\ref{eq:modulation_depth}), $\alpha$
is the relative phase between two interfering laser lights.
For this purpose the piezoelectric stage (P-752.21C, PI) which has 0.2 nm resolution
is installed in the one of the laser light path. By moving the stage, the path length of one
laser light change and hence the relative phase between two laser lights is changed.
Further information can be found in \cite{suehara_NIM}.

\subsection{Gamma-ray Detector}
\begin{wrapfigure}{r}{0.5\columnwidth}
%\begin{figure}
\centerline{\includegraphics[width=0.48\columnwidth]{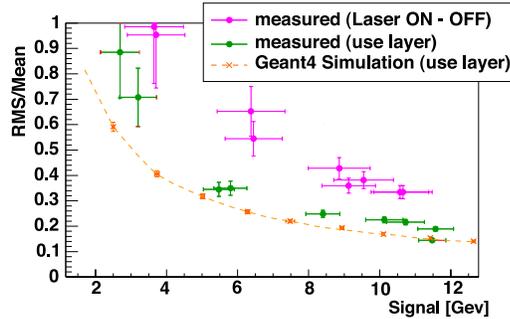}}
\caption{comparison of the measured signal resolution when using multi layer information or not
and the Geant4 simulation}
\label{fig:signal_resolution}
\end{wrapfigure}
At the ATF2, the electron beam energy is 1.3 GeV. Then the average energy of gamma rays
from the inverse Compton scattering between a electron in the beam and 532 nm wavelength
laser photon is 15 MeV on average. The background gamma rays mainly come from the bremsstrahlung 
of the electron beam halo colliding the wall of beam pipe. This background gamma rays have
broad spectrum from 1.3 GeV to lower energy.

It is challenge to measure the signal gamma rays in such a high energy background condition.
To solve this issue the multi-layer electromagnetic calorimeter is developed\cite{oroku_IEEE} which
uses CsI(Tl) as a scintillator. 
%Since lower energy gamma ray deposit energy in forward of a 
%calorimeter, by using multi layered calorimeter the amount of lower energy signal gamma rays and higher energy
%bakcground gamma rays are obtained respectively. 
Based on the fact that the shower depth in the calorimeter is shallower for lower energy signal gamma rays,
multi-layer calorimeter is able to distinguish between lower energy signal and the higher energy background 
by measuring the shower development in the calorimeter. By using the energy deposit in each layer,
the improvement of the signal resolution is observed. Figure \ref{fig:signal_resolution} shows the comparison 
of the signal resolution obtained by two methods. "Laser ON - OFF" means the total energy deposit
in the calorimeter when the laser is on minus when the laser is off, "user layer" means the signal obtained
by the above method.

\section{Measurement}
%% section headers !

\subsection{Procedure}
Beam size measurement is performed as follows.
\begin{enumerate}
\item Adjust the laser pulse and the electron beam timing
\item Scan the laser beam position perpendicular to the electron beam axis and find 
the position where the scattered gamma-ray signal become the maximum
\item Move one laser beam position in the electron beam axis in small step and 
move the laser phase to find the position where two laser lights overlap.
\item Repeat the these steps until the measured modulation depth reaches the maximum
\end{enumerate}
The current beam size is obtained from the maximum modulation depth after these
procedures. If the electron beam condition is changed, repeat these procedures 
from the top (if needed).

\begin{wrapfigure}{r}{0.5\columnwidth}
%\begin{figure}
\centerline{\includegraphics[width=0.45\columnwidth]{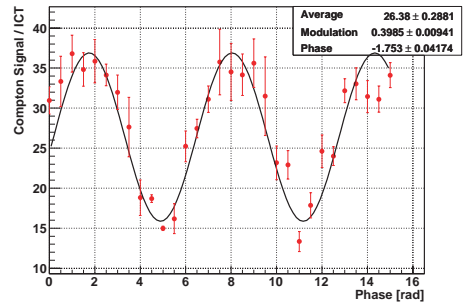}}
\caption{one of the beam size measurement plot}
\label{fig:modulation_plot}
\end{wrapfigure}
\subsection{Result}
Measured beam size so far is several microns with larger $\beta$ optics than
the ATF2 base design (Table \ref{tab:beam_param}).

Figure \ref{fig:modulation_plot} shows one of the plots of smallest beam size measurement.
Measured plots become sinusoidal because the pattern of the laser
fringe pitch can be written as Equation (\ref{eq:laser_dist}). The modulation depth
is obtained by fitting this plot by a sinusoidal function and it was 0.40 in this measurement.
The laser crossing angle was 2.7 degrees
(in this case the fringe pitch was calculated to 11 $\mu$m). Then the beam size
is calculated to about 2.4 $\mu$m by using Equation (\ref{eq:modulation_depth}).

%\begin{wraptable}{l}{0.5\columnwidth}
\begin{table}[htbp]
\centerline{\begin{tabular}{|c|c|c|}
\hline
parameter & present & design \\ \hline \hline
horizontal $\beta$ function & 4 cm &  4 mm\\ \hline
vertical $\beta$ function & 1 mm & 0.1 mm \\ \hline
horizontal beam size & 10 $\mu$m(measured) & 2.8 $\mu$m \\ \hline
vertical beam size & 2.4 $\mu$m(measured) & 37 nm \\ \hline
\end{tabular}}
\caption{parameters of the electron beam at the interaction point of ATF2}
\label{tab:beam_param}
%\end{wraptable}
\end{table}

%\begin{wrapfigure}{r}{0.5\columnwidth}
\begin{figure}[htbp]
\centerline{\includegraphics[width=0.45\columnwidth]{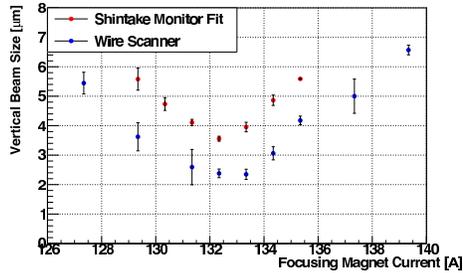}}
\caption{comparison of the beam size measurements of Shintake monitor and
the wire scanner}
\label{fig:qscan_compare}
%\end{wrapfigure}
\end{figure}

\subsection{Systematic Error}

During the beam size measurement using Shintake monitor, the other conventional
beam size monitor was also used to evaluate the systematic error of Shintake monitor.
For this purpose a wire scanner (which is made of 10 $\mu$m diameter tungsten wire) was
installed at the same place. Its resolution limit is about 2.5 $\mu$m.

Figure \ref{fig:qscan_compare} shows the 
comparison of the measurements of Shintake monitor and the wire scanner obtained 
by changing the strength of the final focusing magnets in the end of 2009.
 Clearly the systematic error exists between them.
This is closely discussed in the presentation by Y.~Yamaguchi \cite{yamaguchi_slide}.

\section{Summary and Schedule}
Shintake monitor at ATF2 is now ready for measurement and succeeded in the observation
of the gamma-ray modulation from the inverse Compton scattering between the electron
beam and the laser light. The measured vertical beam size is several microns because ATF2
runs with the large $\beta$ optics for the moment.

The current beam time will be continued until this June and smaller $\beta$ optics will
be tested during this period. After the summer shutdown the optics will be shifted to the designed one and 
then, 
tuning of the accelerator and the measurement of 37 nm vertical beam size will be tried.

% ****************************************************************************
% BIBLIOGRAPHY AREA
% ****************************************************************************

\begin{footnotesize}
% IF YOU DO NOT USE BIBTEX, USE THE FOLLOWING SAMPLE SCHEME FOR THE REFERENCES
% ----------------------------------------------------------------------------

% ----------------------------------------------------------------------------

\end{footnotesize}

% ****************************************************************************
% END OF BIBLIOGRAPHY AREA
% ****************************************************************************

\end{document}